\documentclass{ws-ijmpa}
\usepackage{subfigure}
\begin{document}

\markboth{L. Le\'sniak, A. Furman, R. Kami\'nski, B. El-Bennich, B. Loiseau}
{Final state interactions in $B\to \pi\pi K$ and 
$B\to K\overline KK$ decays}

%
\catchline{}{}{}{}{}
%

\title{\bf FINAL STATE INTERACTIONS IN $B\to \pi\pi K$ and 
$B\to K\overline KK$ DECAYS}

\author{\footnotesize L. LE\'SNIAK$^1$, B. EL-BENNICH$^2$, A. FURMAN$^3$,
 R. KAMI\'NSKI$^1$, B. LOISEAU$^2$}
\vspace{0.5cm}
\address{\it \small  $^1$ Division of Theoretical Physics, The Henryk Niewodnicza\'nski 
Institute of Nuclear Physics, Polish Academy of Sciences, 31-342 Krak\'ow,
 Poland \\ Leonard.Lesniak@ifj.edu.pl\\
  $^2$Laboratoire de Physique Nucl\'eaire et de Hautes 
\'Energies (IN2P3-CNRS-Universit\'es Paris 6 et 7), Groupe Th\'eorie,\\
Universit\'e Pierre et Marie Curie, 4 Pl. Jussieu, F-75252 Paris, France\\
$^3$ ul. Bronowicka 85/26, 30-091 Krak\'ow, Poland} 
 


\maketitle


\begin{abstract}
Analysis of charged and neutral $B$ meson decays into
$\pi^+\pi^-K$, $K^+K^-K$ and $K^0_SK^0_SK^0_S$ is performed
using a unitary representation of the $\pi\pi$ and $K\overline K$ final state interactions.
Comparison of the theoretical model with the 
experimental data of the Belle and BaBar Collaborations
indicates that charming penguin contributions are necessary to describe the
 $B \to f_0(980) K$ and $B \to \rho(770)^0 K$ decays. 

\keywords{Decays of bottom mesons; meson-meson interactions}

\end{abstract}

\ccode{PACS numbers: 13.25.Hw, 13.75Lb}

\section{\bf Introduction}

We report on some studies of three-body $B$ meson decays into $\pi\pi K$ and 
$ K\overline KK$ final states. In these reactions one can find an
evidence of the direct $CP$ violation similar to that recently discovered in two-body
$B^0$ decays into $\pi^{\pm}K^{\mp}$. Loop-type weak decay diagrams known as
penguin terms play an important role in these decays. In particular we study
contributions of the charming penguin amplitudes responsible for 
long-distance effects present in the final state interactions. Strong
interactions between pairs of pions and kaons in the final scalar-isoscalar state have
been described in Ref.~ \refcite{fkll}. Here we extend this approach to the $\pi\pi$
interactions in $P$-wave. This enables us to obtain a unitarized
description of the final state $\pi\pi$ interactions from  threshold
up to about 1.2 GeV. The opening of the $K \overline  K$ threshold near 1 GeV is
included in a natural way since both $\pi\pi$ and 
$ K\overline K$  $S$-wave channels are coupled in our model.

\section{\bf Decay amplitudes}

We take into account two components in the weak transition amplitudes governed by 
$b$-quark decays into $s\overline u u$, $s\overline d d$ and $s\overline s s$,
where $u$, $d$ and $s$ denote up, down and strange quarks, respectively.
The first term consists of the  amplitude derived in the factorization 
approximation with some QCD corrections and the second one is a long-distance
amplitude with $c$-quark or with $u$-quark in loop. At the hadronic level
the second amplitude with the $c$-quark in loop can be interpreted in terms of the 
intermediate $D_s^{(*)}D^{(*)}$ states which are frequently produced in
$B$-meson decays (see Fig. 1).
\begin{center}
\begin{figure}[h!]
\includegraphics*{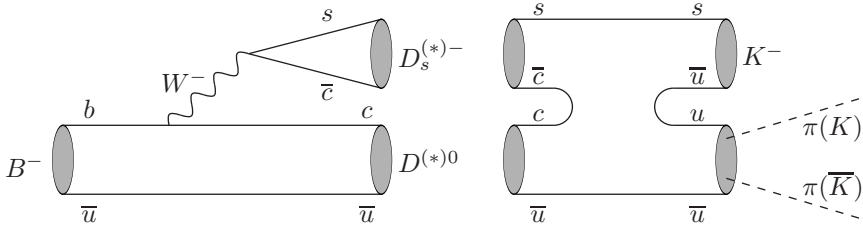}
\caption{Example of $D_s^{(*)}D^{(*)}$ contribution to $B^-$ decays into
 $\pi\pi K^-$ and $K\overline KK^-$}
\label{fig1}
\end{figure}
\end{center}   
The charming penguin contribution to the $B^-\to \pi\pi K^-$ decay amplitude with
two pions in  $P$-wave is parametrized by the following expression:
\begin{equation}
\langle(\pi^+\pi^-)_{P}K^-\vert H\vert B^-\rangle_{penguin}
=2~ G_F m_{\rho} C_{\rho} \Gamma_{\rho\pi\pi}(m_{\pi\pi}) p_{\pi}p_{K} cos\theta,
\label{Penguin}
\end{equation}
where $G_F$ is the Fermi constant, $m_{\rho}$ is the $\rho(770)$ mass,
$\Gamma_{\rho\pi\pi}(m_{\pi\pi})$ is the $\rho\pi\pi$ vertex function of the
$\pi\pi$ effective mass $m_{\pi\pi}$, $p_{\pi}$
and $p_{K}$ are the pion and kaon momenta in the $\rho$ rest frame and $\theta$
is the angle between the direction of flight of the $\pi^-$ and the direction
of the $\pi^+\pi^-$ system in the $B$ rest frame. The constant amplitude 
$C_{\rho}$ reads:  
\begin{equation}
C_{\rho}=f_K\,A_0^{B\to\rho}(m_K^2)\left(V_{ub}V_{us}^*P_u+V_{tb}V_{ts}^*P_t\right).
\label{cro}
\end{equation}
Here $V$'s are the  Cabibbo--Kobayashi--Maskawa matrix elements, $f_K$ is the 
kaon decay constant, $A_0^{B\to\rho}(m_K^2)$ is the $B \to \rho$ transition form
factor and 
$P_u$ and  $P_t$ are penguin $P$-wave complex parameters to be fitted to the experimental data.

Final state interactions in the isospin zero $S$-wave are treated using the unitary model
 of Ref.~ \refcite{KLM} in which the $\pi\pi$ and $K \overline K$ channels
are coupled. In this approach the sum of the several Breit-Wigner terms, usually
 used in the 
experimental analyses of the Dalitz plot distributions, is replaced by
 a set of unitary coupled meson-meson amplitudes.
These amplitudes are expressed in terms of phase shifts 
{$\delta_{\pi\pi}$, $\delta_{KK}$} and  inelasticity $\eta$ 
known from other experiments \cite{KLM}. The $P$-wave pion-pion amplitude is well described by a Breit-Wigner 
term.
No arbitrary phases nor relative intensity free parameters for the different
resonances are needed. All the resonances appear in a natural way as poles of 
the meson-meson amplitudes. The scalar resonances $f_0(600)$ and $f_0(980)$  are examples 
of such poles of a single amplitude. The presence of the $f_0(980)$ is a dominant
feature of the experimental $m_{\pi\pi}$ distribution.  

\section{\bf Comparison with the experimental data}

We  perform a fit to the data of Belle \cite{Belle1,Belle2} and BaBar
\cite{BaBar} Collaborations. Both groups have measured differential
  $\pi\pi$ effective mass
distributions, branching fractions, direct $CP$ violating asymmetries in charged
$B$ decays and time-dependent $CP$ violating asymmetry parameters in $B^0$ decays.
Moreover, the Belle Collaboration has given helicity angle distributions
in the $\rho(770)$ and $f_0(980)$ regions. We obtain a good agreement
with the data using only the four complex penguin parameters:
$S_u = (0.15 \pm 0.10) \exp [i(1.90 \pm 0.71)]$, $S_t = (0.020 \pm 0.002)
\exp [-i(0.26 \pm 0.21)]$ for the $S$-wave and $P_u = (1.09 \pm 0.21)
 \exp [-i(0.98 \pm 0.12)]$, $P_t = (0.065 \pm 0.002)\exp [-i(1.56 \pm 0.08)]$ 
for the $P$-wave. 
\begin{figure*}[h!]
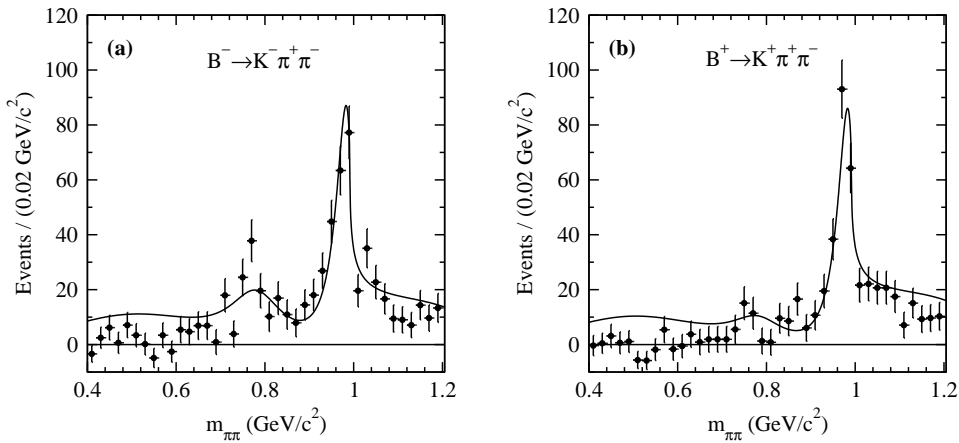

\subfigure{\includegraphics*[height=0.3\textheight]{fig2a.eps}}~~~~~
\subfigure{\includegraphics*[height=0.3\textheight]{fig2b.eps}}
\caption{The $\pi^+\pi^-$ effective-mass distributions in
 $B^{\pm} \to \pi^+\pi^- K^{\pm}$
 decays. The data points are taken from Ref. 4 and the 
 solid lines represent the results of our model.} 
\label{fig2}
\end{figure*}

In Fig. 2 we show a comparison of our model with the $\pi^+\pi^-$ effective mass
distributions measured by Belle for the $B^+\to \pi\pi K^+$ and
$B^-\to \pi\pi K^-$ decays. The large direct $CP$ asymmetry of the order of 0.3,
visible in the range of the $\rho(770)$ resonance, is well described  by our 
model. The fit is performed in the $m_{\pi\pi}$ range between 0.60 GeV and 1.06
GeV where {the two resonances $\rho(770)$ and $f_0(980)$ dominate the pion-pion mass
spectrum.
Existence of these resonances leads to interesting interference phenomena
seen in Fig. 3 where  helicity-angle distributions are plotted for the
combined $B^{\pm}\to \pi\pi K^{\pm}$ decays. In the $\rho$ range the general
behaviour of the data follows the $\cos ^2\theta$ function, characteristic
of a polarized $\rho$ decay into $\pi^+\pi^-$. However, in the $f_0(980)$ 
range the angular dependence is not flat as one can expect for the decay of a 
$S$-wave resonance. It has an interference component proportional to 
$\cos \theta$ which
originates from the presence of the $\rho$ resonance tail under the 
$f_0(980)$ peak.
\begin{center}
\begin{figure*}[h!]
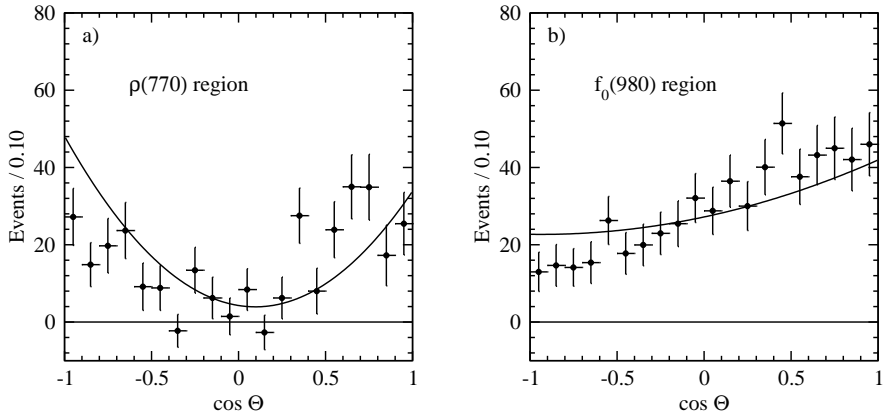

\subfigure{\includegraphics*[height=0.28\textheight]{fig3a.eps}}~~~~~
\subfigure{\includegraphics*[height=0.28\textheight]{fig3b.eps}}
\caption{Angular distributions in $B^\pm \to \pi^+\pi^- K^\pm$ {\bf (a)} in the
$\rho(770)$ mass region and {\bf (b)} in the $f_0(980)$ region.
The data points are from Ref. 4 and the solid lines denote our model.}
\label{fig3}
\end{figure*}
\end{center}   

We have also performed calculations of different observables for the $B$ decays
into three kaons in which a $K \overline K$ pair is in  relative $S$-wave.
No extra free parameters have been used since all the parameters have been
fixed in the fit to the $B\to \pi\pi K$ decays as described above. The averaged 
branching ratio for the $B^\pm \to K^+K^- K^\pm$ 
$S$-wave channels, integrated over the 
$K \overline K$ effective masses from the threshold
up to 1.1 GeV, is equal to
 $1.7 \times 10^{-6}$ 
 and the corresponding value for the 
$B^0\to K^+K^- K^0$ is
  $0.9 \times 10^{-6}$. The direct $CP$ asymmetry for the charged $B$
decays is 
  $ 0.07$. The time-dependent asymmetry parameters for the neutral
$B$ decays are:
$\mathcal S = -0.80 $ and $\mathcal A =- 0.13$.

\section*{Acknowledgments}

This work has been performed in the framework of the IN2P3-Polish 
Laboratories Convention (project No. CSI-12). It has been also
 financed within an agreement between the CNRS (France) and the Polish Academy
  of Sciences (project No.~19481).


\begin{thebibliography}{0}    


\bibitem{fkll} A. Furman, R. Kami\'nski, L. Le\'sniak and B. Loiseau,
{\it Phys. Lett.} {\bf B622}, 207 (2005).

\bibitem{KLM} R.~Kami\'nski, L.~Le\'sniak and B. Loiseau, 
{\it Phys. Lett.} {\bf B413}, 130 (1997).

\bibitem{Belle1} Belle Collaboration (A. Garmash {\it et al.}), {\it hep-ex}/0512066.

\bibitem{Belle2} Belle Collaboration (K. Abe {\it et al.}), {\it hep-ex}/0509001.

\bibitem{BaBar} BaBar Collaboration (B. Aubert {\it et al.}), {\it Phys. Rev.}
 {\bf D72}, 072003 (2005).

\end{thebibliography}
\end{document}